# Hidden progress: broadband plasmonic invisibility

Jan Renger[1+], Muamer Kadic[2+], Guillaume Dupont[2], Srdjan S. Aćimović[1], Sébastien Guenneau[2], Romain Quidant[1], Stefan Enoch,[2*]

[1] *ICFO-Institut de Ciencies Fotoniques, Mediterranean Technology Park,*

*08860 Castelldefels (Barcelona), Spain*

[2] *Institut Fresnel, CNRS, Aix-Marseille Université,*

*Campus Universitaire de Saint-Jérôme, 13013 Marseille, France*

[+] *'These authors contributed equally to this work'*

**Abstract:**

**The key challenge in current research into electromagnetic cloaking is to achieve invisibility over an extended bandwidth. There has been significant progress towards this using the idea of cloaking by sweeping under the carpet of Li and Pendry, with dielectric structures superposed on a mirror. Here, we show that we can harness surface plasmon polaritons at a metal surface structured with a dielectric material to obtain a unique control of their propagation. We exploit this to control plasmonic coupling and demonstrate both theoretically and experimentally cloaking over an unprecedented bandwidth (650-900 nm). Our non-resonant plasmonic metamaterial allows a curved reflector to mimic a flat mirror. Our theoretical predictions are validated by experiments mapping the surface light intensity at the wavelength 800 nm.**

**Introduction**

In 2006, Pendry et al. [1] and Leonhardt [2] independently showed the possibility of designing a cloak that renders any object inside it invisible to electromagnetic radiation. This coating consists of a meta-material whose physical properties (permittivity and permeability) are deduced from a coordinate transformation in the Maxwell system. The anisotropy and the heterogeneity of the parameters of the coat work as a deformation of the optical space around the object. The first experimental validation [3] of these theoretical considerations was given, a few months later for a copper cylinder invisible to an incident plane wave at 8.5 GHz as predicted by the numerical simulations. This markedly enhances our capabilities to manipulate light, even in the extreme near field limit [4]. However, such cloaks suffer from an inherent narrow bandwidth as their transformation optics design leads to singular tensors on the frontier of the invisibility region (obtained by blowing up a point [5]). To remove the cloak's singularity, Xiang et al. proposed to consider the blow-up of a segment instead of a point [6], but this cloak only works for certain directions. On the other hand, Leonhardt and Tyc considered a stereographic projection of a virtual hyper-sphere in a four dimensional space [7], which bears some resemblance with the construction of a Maxwell fisheye.

As an alternative to non-singular cloaking Li and Pendry proposed a one-to-one geometric transform from a flat to a curved ground: their invisibility carpet [8] is by essence non singular and thus broadband. This proposal led to a rapid experimental progress in the construction of carpets getting close to optical frequencies [9-11].

Another way to make cloaks broadband is to approximate their parameters using a homogenization approach, which leads to nearly ideal cloaking [12, 13, 14], as it does not rely upon locally resonant elements. In 2008, some of us [15] demonstrated broadband cloaking of surface liquid waves using a micro-structured metallic cloak which was experimentally validated around 10 Hz. Ultra-broadband cloaking can even be achieved in this way for flexural surface waves in thin-plates [16]. This naturally prompts the question of whether, at optical frequencies, an object lying onto a metal film could be cloaked from propagating Surface Plasmon Polaritons (SPPs).

The physics of the extraordinary transmission of light through holes small compared with the wavelength is by now well known [*17*], but some heralding earlier work is less well known (for example combining both theory and experiment [18]. Pendry, Martin-Moreno and Garcia-Vidal further showed in 2004 that one can manipulate surface plasmon ad libitum via homogenization of structured surfaces [19]. In the same vein, pioneering approaches to invisibility relying upon plasmonic metamaterials have already led to fascinating results [20-23]. These include plasmonic shells with a suitable out-of-phase polarizability in order to compensate the scattering from the knowledge of the electromagnetic parameters of the object to be hidden, and external cloaking, whereby a plasmonic resonance cancels the external field at the location of a set of electric dipoles. Recently, Baumeier et al. have demonstrated theoretically and experimentally that it is possible to reduce significantly the scattering of an object when it is surrounded by two concentric rings of point scatterers [23].

In the present letter, we extend the design of invisibility carpets to SPPs. The cornerstone of our approach is the identification of the dispersion relation of SPPs at the interface between a metal and a heterogeneous anisotropic carpet. On this basis, we design a dielectric carpet and use full wave computations to study its properties. Finally, we measure the SPP intensity when interacting with the carpet to validate the concept.

**Transformation plasmonics for invisibility carpets**

We consider a transverse magnetic ($p$-polarized) SPP propagating in the positive $x$ direction at the interface $z = 0$ between metal ($z < 0$) and air ($z > 0$):

$$\begin{cases} \mathbf{H}_2 = (0, H_{y2}, 0) \exp\{\imath(k_{x2}x - \omega t) - k_{z2}z\},\ z > 0, \\ \mathbf{H}_1 = (0, H_{y1}, 0) \exp\{\imath(k_{x1}x - \omega t) + k_{z1}z\},\ z < 0, \end{cases} \quad (1)$$

with $\Re(k_{z1})$ and $\Re(k_{z2})$ being strictly positive in order to maintain evanescent fields above and below the interface $z = 0$.

For this field to be solution of Maxwell's equations, continuity of its tangential components is required across the interface $z = 0$ and this brings $k_{x1} = k_{x2} = k_x$ together with the dispersion relations

$$k_{zi} = \sqrt{k_x^2 - \varepsilon_i \left(\frac{\omega}{c}\right)^2},\ \frac{k_{z1}}{\varepsilon_1} + \frac{k_{z2}}{\varepsilon_2} = 0, \quad (2)$$

where $c$ is the speed of light in vacuum, $\varepsilon_1(x, y) = 1$ in air and $\varepsilon_1(x, y) = 5.76$ in the pillars ($z > 0$), and $\varepsilon_2 = 1 - \frac{\omega_p^2}{\omega^2 + i\gamma\omega}$, the usual Drude form in the metal ($z < 0$), for which $\omega_p$ is the plasma frequency (2175 THz) of the free electron gas and $\gamma$ is a characteristic collision frequency of about 4.35 THz [24]. Altogether, the condition

$$k_x = \frac{\omega}{c} \sqrt{\frac{\varepsilon_1 \varepsilon_2}{\varepsilon_1 + \varepsilon_2}}, \quad (3)$$

should be met for SPPs to be able to propagate at the interface. SPPs are bound to the interface, hence, do not belong to the radiative spectrum (unlike leaky waves). We note that since $k_{z1}$ and $k_{z2}$ are strictly positive (remove and) SPPs can only exist if $\varepsilon_1$ and $\varepsilon_2$ are of opposite signs.

So far, such a mathematical setting is fairly standard. However, we now wish to analyse the interaction of SPPs with an anisotropic heterogeneous structure, in the present case an invisibility carpet, deduced from the following geometric transformation:

$$\begin{cases} x' = \dfrac{x_2(y) - x_1(y)}{x_2(y)} x + x_1(y), \, 0 < x < x_2(y), \\ y' = a < y < b, \\ z' = z, \, 0 < z < +\infty, \end{cases} \quad (4)$$

where $x'$ is a stretched vertical coordinate. It is easily seen that this linear geometric transform maps the segment $(a,b)$ of the horizontal axis $x = 0$ onto the curve $x' = x_1(y)$, and it leaves the curve $x = x_2(y)$ unchanged. The curves $x_1$ and $x_2$ are assumed to be differentiable, and this ensures that the carpet won't display any singularity on its inner boundary.

The permittivity and permeability tensors in the transformed coordinates are now given by:

$$\underline{\underline{\varepsilon}}' = \varepsilon_0 \mathbf{T}^{-1}, \quad \text{and} \quad \underline{\underline{\mu}}' = \mu_0 \mathbf{T}^{-1} \text{ where } \mathbf{T} = \mathbf{J}^T \mathbf{J} / \det(\mathbf{J}), \quad (5)$$

with $\mathbf{J}$ the Jacobian matrix of the transformation and $\varepsilon_0 \mu_0 = 1/c^2$.

The symmetric tensors $\underline{\underline{\varepsilon}}'$ and $\underline{\underline{\mu}}'$ are fully described by five non vanishing entries in a Cartesian basis:

$$(T^{-1})_{11} = \left(1 + \left(\dfrac{\partial x}{\partial y'}\right)^2\right)\alpha, \, (T^{-1})_{12} = (T^{-1})_{21} = -\dfrac{\partial x}{\partial y'}$$

$$(T^{-1})_{22} = \dfrac{1}{\alpha}, \, (T^{-1})_{33} = \dfrac{1}{\alpha}, \quad (6)$$

where $\alpha = (x_2 - x_1)/x_2$.

In a diagonal basis associated with a quasi-conformal grid the fully anisotropic tensors reduce to: $\underline{\underline{\varepsilon}}' = \mathrm{diag}(\varepsilon_{xx2}, \varepsilon_{yy2}, \varepsilon_{zz2})$ and $\underline{\underline{\mu}}' = \mathrm{diag}(\mu_{xx2}, \mu_{yy2}, \mu_{zz2})$ as described in the supporting online material. Further assuming some low spatial variation of both tensors within the carpet, we define local transverse wave numbers associated with SPPs propagating at the interface between metal and the carpet that should satisfy (2) for $i = 1$. They should however satisfy the following equations (27) for $i = 1, 2$:

$$k_{zi} = \sqrt{\varepsilon_{xx2}\left(\frac{k_x^2}{\varepsilon_{zz2}} - \mu_{yy2}\left(\frac{\omega}{c}\right)^2\right)}, \quad \frac{k_{z1}}{\varepsilon_1} + \frac{k_{z2}}{\varepsilon_{xx2}} = 0. \tag{7}$$

The dispersion relation (3) for the surface polaritons takes the following form

$$k_x = \frac{\omega}{c}\sqrt{\frac{\varepsilon_{zz2}\varepsilon_1(\mu_{yy2}\varepsilon_1 - \varepsilon_{xx2})}{\varepsilon_1^2 - \varepsilon_{xx2}\varepsilon_{zz2}}}. \tag{8}$$

Note that if we assume that $\varepsilon_{xx2} = \varepsilon_{zz2} = \varepsilon_2$, and $\mu_{yy2} = 1$, we retrieve (3) as we should.

We note that there exists a family of coordinate transformations that would map a flat mirror onto a curved one. However, if we now want to reduce the anisotropy of the carpet (e.g. to avoid dealing with magnetism), we need to work with quasi-conformal coordinates, as proposed by Li and Pendry [8]. We deduced those from the minimization of the Modified Liao functional [26], and the resulting grid is shown in Fig.1a.

In Fig.1b we have simulated a SPP incident on a heterogeneous carpet whose refractive index distribution is related to the quasi-conformal map via $n^2 = 1/\sqrt{\det(\mathbf{J}^T\mathbf{J})}$. We are in a position to mimic the properties of such a spatially varying isotropic metamaterial simply by placing some identical dielectric pillars at the nodes of the quasi-conformal grid, as shown in Fig.1a. The diameter of these pillars has been found via a simple optimization algorithm with an objective function taking as argument the refractive index of pillars and their diameter.

Moreover, it is easily seen that the anisotropy of the crescent carpet reduces when we flatten its boundaries, and this allows us to find a good compromise between control of the SPPs and complexity of the arrangement of pillars. We discovered that dielectric cylindrical pillars with a refractive index $n = \sqrt{5.76}$, of height 200 nm and identical diameter 200 nm lead to nearly perfect cloaking when they are located at the quasi-conformal grid nodes.

We report the result of our computations for a SPP at $\lambda$ = 800 nm incident upon an array of cylindrical dielectric pillars, in Fig. 2(d) and for comparison that of a metallic bump in Fig. 2(c). The reduced backscattering is clearly visible. Importantly, we further numerically checked that the SPP cloaking is robust with respect to the shape of pillars, and for instance conical pillars produce the same overall scattering as cylindrical pillars of the same height (not shown for sake of brevity). We attribute this property to the fact that the dielectric pillars are non-resonant elements whose scattering cross-section is only slightly modified by moderate changes in shape.

**Experimental results**

In order to meet experimentally the parameters found in our simulations, we chose a configuration in which a gold surface is structured with $TiO_2$ nanostructures. The $TiO_2$ pillars forming the crescent-moon-like carpet were first fabricated on top of a 60-nm-thin Au film by combining electron-beam lithography and reactive ion etching. In a second lithography step, we added a curved Bragg-type reflector (formed by 15 gold lines (section = 150 nm × 150 nm) periodically separated by half the SPP wavelength), acting as the object to be hidden behind the carpet (see Fig. 1(c)). The shape of the obtained TiO$_2$ particles is conical (h = 200 nm, r = 210 nm) as a consequence of the etching anisotropy.

The SPP was launched at a ripple-like, 200-nm-wide TiO$_2$-line placed 44 $\mu$m away from the reflector. SPPs propagating on thin metal films deposited on dielectric substrate have radiative losses into the substrates. This leakage radiation was collected using a high-numerical aperture objective to map the SPP fields [24]. Additionally for the sake of clarity, we employed spatial filtering in the conjugated (Fourier-) plane to suppress the direct transmitted light from the excitation spot and scattered light in order to isolate the carpet properties.

The leakage radiation microscopy (LRM) images map the distribution of the (replace SPP by SPPs) propagating at the gold/air interface and interacting with the different structures fabricated at the gold surface. In the case of a bare curved Bragg-reflector, the reflected SPPs are propagating into different directions depending on their relative angle to the normal to the mirror lines (see green arrows in Fig. 3 (c)), thus leading to a curved wave front. Conversely, adding the crescent-moon-like TiO$_2$ carpet re-establishes a fringe pattern with a nearly straight wave front (see Fig. 3 (b)) very similar to the case of a flat Bragg-mirror. The remaining small lateral modulations are attributed to imperfections in the manufacturing.

**Conclusion**

In conclusion, we have studied analytically and numerically the extension of cloaking to near infrared SPP waves propagating at a metal/dielectric interface. These waves obey the Maxwell equations at a flat interface and are evanescent in the transverse direction, so that, the problem we have treated is somewhat isomorphic to the case of linear surface water waves which satisfy a similar dispersion relation.

Nevertheless, the presented analytical derivation of the dispersion relation for SPPs propagating at the interface between metal and transformation-based anisotropic medium bridges transformation optics and plasmonics. Moreover, numerical computations based on the finite element method take into account the three dimensional features of the problem, such as plasmon polarization and jump of permittivity at the interface between metal and a plasmonic carpet consisting either of an anisotropic heterogeneous medium or dielectric pillars regularly spaced in air. These two media are in any case described by permittivities of opposite sign to leave enough room for the existence of SPPs.

One of the main achievements of this letter is to bring cloaking a step closer to visible wavelengths as we consider a SPP at $\lambda = 800$ nm. We also emphasize that the manufactured crescent carpet should be broadband. Numerical results shown in the supporting online material for a range of wavelengths $\lambda$ = 650 to 900 nm demonstrate the principle would also work in the visible spectrum. However, SPPs attenuate faster at such wavelengths which pose future experimental challenges. Finally, work is in progress to engineer an invisibility cloak along the same lines.

The rapid experimental progress in the construction of carpets getting close to optical frequencies [10,11,9], as well as recent proposals of metamaterials emulating anisotropic permittivity and permeability [26] suggest that our design might foster efforts in these directions.

J. R., S.S. A., and R. Q. acknowledge the support from La Fundació CELLEX Barcelona. M. K. and G. D. are thankful for the PhD scholarship from the French Ministry of Higher Education and Research. S.

G. acknowledges funding from the Engineering and Physical Sciences Research Council grant EPF/027125/1. The authors are thankful for insightful comments from R.C. McPhedran.

'**Supplementary Information** accompanies the paper on **www.nature.com/nature**.'

* To whom correspondence should be addressed; E-mail: stefan.enoch@fresnel.fr

**Figures**

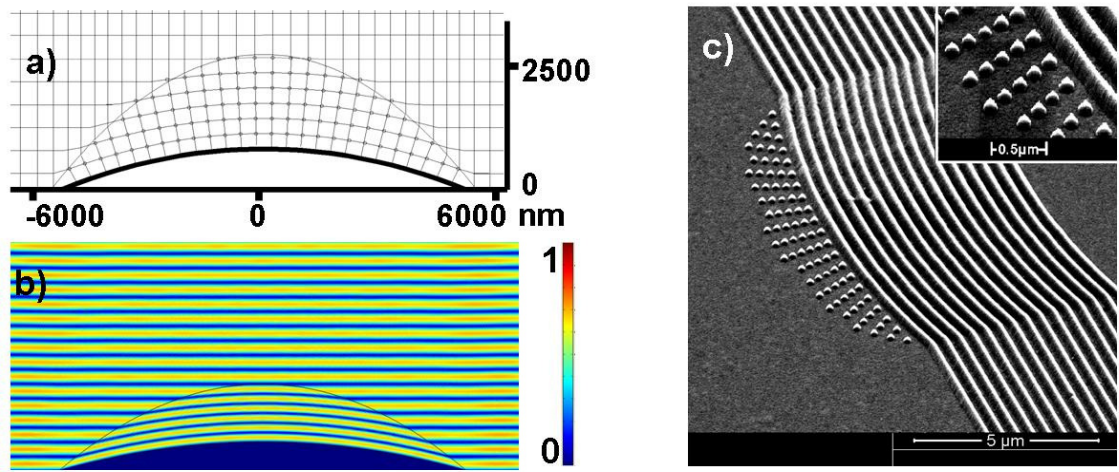

Figure 1. (a) Quasi-conformal grid associated with transformed plasmonic space in a crescent carpet; (b) SPP incident from the top on the heterogeneous carpet deduced from the quasi-conformal mapping (c) SEM micrograph of the structure realized by single-step ebeam-lithography. The cloak is made of $TiO_2$ cones as shown in the zoom (upper right). The $TiO_2$ particles have a conical shape numerically approximated by a cylindrical one (h=200 nm, r=210 nm).

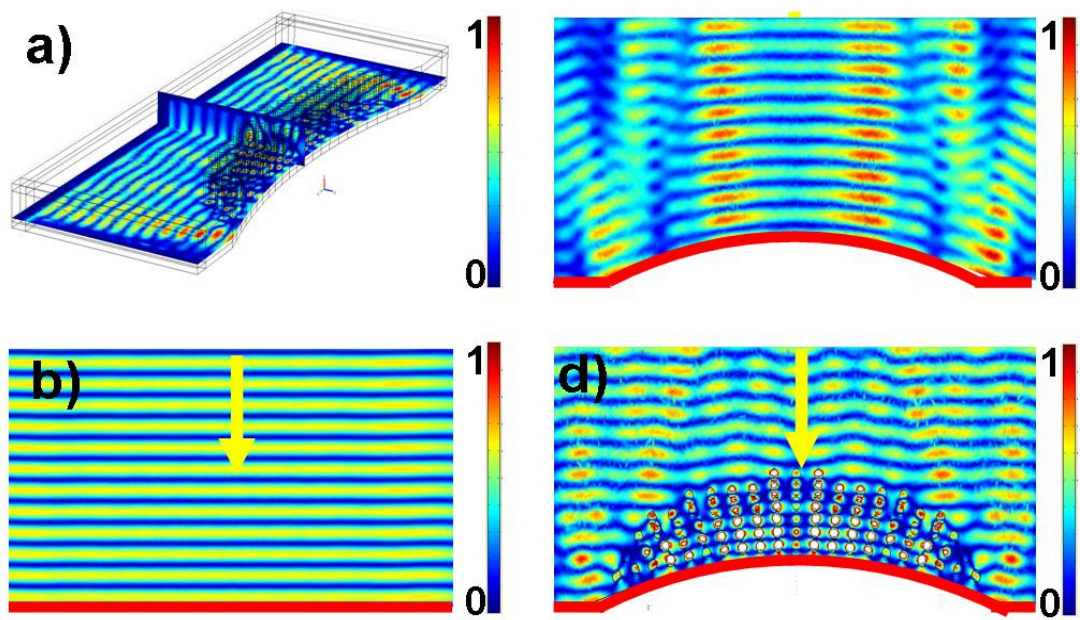

Figure 2. Numerical diffraction of a SPP incident from the top for (b), (c) and (d). Magnitude of the magnetic field is represented: (a) 3D structure used in the simulations.(b) The incident SPP (yellow arrow) hits the straight reflector (red line). (c) The incident SPP (yellow arrow) hits the curved reflector (red line). (d) Cloak is placed in front of the curved reflector (red line). The Beating pattern in backreflection is clearly visible and similar to a straight reflector.

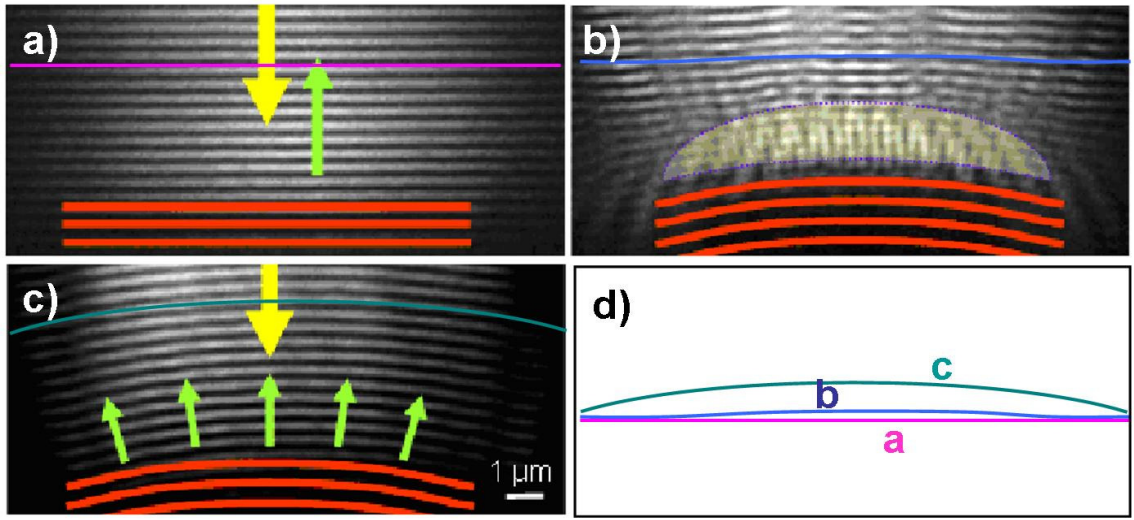

Figure 3. Experimental image of the leakage radiation of the SPP at λ=800 nm. (a) The incident SPP (yellow arrow) hits the straight Bragg mirror (red). (b) The incident SPP (yellow arrow) hits the curved Bragg mirror (red). The backreflected SPPs (green) have different directions due to the curved shape of the reflector resulting in the curved intensity pattern dominated by the beating of the counterpropagating SPPs. (c) Cloak is placed in front of the curved Bragg mirror. The Beating pattern in backreflection is clearly visible and similar to a straight Bragg reflector. (d) Averaged relative position of the interference fringes

**Supplementary material**

**Surface plasmons at the interface between metal and transformed medium with anisotropic permittivity and permeability**

Let us derive the dispersion relation for a surface plasmon at the interface between metal and the above anisotropic media described by diagonal tensors of relative permittivity and permeability $\underline{\underline{\varepsilon'}} = \mathrm{diag}(\varepsilon_{xx2}, \varepsilon_{yy2}, \varepsilon_{zz2})$ and $\underline{\underline{\mu'}} = \mathrm{diag}(\mu_{xx2}, \mu_{yy2}, \mu_{zz2})$. From the first Maxwell equation, we know that

$$\begin{cases} \nabla \times \mathbf{H}_2 = -i\omega\varepsilon_0 \underline{\underline{\varepsilon'}} \mathbf{E}_2, & z>0, \\ \nabla \times \mathbf{H}_1 = -i\omega\varepsilon_0 \varepsilon_1 \mathbf{E}_1, & z<0, \end{cases} \quad (9)$$

where $\mathbf{H}_j$ is defined by:

$$\begin{cases} \mathbf{H}_2 = (0, H_{y_2}, 0)\exp\{\iota(k_{x2}x - \omega t) - k_{z2}z\}, & z>0, \\ \mathbf{H}_1 = (0, H_{y_1}, 0)\exp\{\iota(k_{x1}x - \omega t) + k_{z1}z\}, & z<0, \end{cases} \quad (10)$$

with $\Re(k_{z1})$ and $\Re(k_{z1})$ strictly positive in order to maintain evanescent fields above and below the interface $z=0$. This leads to

$$\begin{cases} \mathbf{E}_2 = -\dfrac{c}{\omega} H_{y_2} \left(\dfrac{k_{z2}}{\varepsilon_{xx2}}, 0, \dfrac{k_{x2}}{\varepsilon_{zz2}}\right)\exp\{\iota(k_x x - \omega t) - k_{z2}z\}, & z>0, \\ \mathbf{E}_1 = -\dfrac{c}{\omega} H_{y_1} \left(\dfrac{k_{z1}}{\varepsilon_1}, 0, \dfrac{k_{x2}}{\varepsilon_1}\right)\exp\{\iota(k_x x - \omega t) - k_{z1}z\}, & z<0, \end{cases} \quad (11)$$

with $\mathbf{E}_j = (E_{xj}, 0, E_{zj})$. The transverse wave numbers are found by invoking the other Maxwell equation

$$\begin{cases} \nabla \times \mathbf{E}_2 = i\omega\mu_0 \underline{\underline{\mu'}} \mathbf{H}_2, & z>0, \\ \nabla \times \mathbf{E}_1 = i\omega\mu_0 \mathbf{H}_1, & z<0, \end{cases} \quad (12)$$

which leads to

$$k_{zi} = \sqrt{\varepsilon_{xx2}\left(\frac{k_x^2}{\varepsilon_{zz2}} - \mu_{yy2}\left(\frac{\omega}{c}\right)^2\right)}, \ j = 1, 2, \qquad (13)$$

The boundary condition at the interface $z = 0$ requires continuity of the tangential components of the electromagnetic field, which is ensured if

$$\frac{k_{z1}}{\varepsilon_1} + \frac{k_{z2}}{\varepsilon_{xx2}} = 0. \qquad (14)$$

Substituting (13) into (14), we obtain the dispersion relation for a surface plasmon at the interface between a metal and an invisibility carpet

$$k_x = \frac{\omega}{c}\sqrt{\frac{\varepsilon_{zz2}\varepsilon_1(\mu_{yy2}\varepsilon_1 - \varepsilon_{xx2})}{\varepsilon_1^2 - \varepsilon_{xx2}\varepsilon_{zz2}}} \ . \qquad (15)$$

## Reduced form of the permittivity and permeability tensors

We can compute the eigenvalues of $\mathbf{T}^{-1}$ as these are the relevant quantities for the tensor components along the main optical axes:

$$\Lambda_i = \frac{1}{2\alpha}\left[1+\alpha^2+\left(\frac{\partial y}{\partial x'}\right)^2\alpha^2+(-1)^{i-1}\sqrt{-4\alpha^2+\left(1+\alpha^2+\left(\frac{\partial y}{\partial x'}\right)^2\alpha^2\right)^2}\right], \Lambda_3 = \frac{1}{\alpha}. \quad (16)$$

with $\alpha = (x_2 - x_1)/x_2$. We note that all eigenvalues $\Lambda_i$ are strictly positive functions as obviously $1+\alpha^2+\left(\frac{\partial y}{\partial x'}\right)^2\alpha^2 > \sqrt{-4\alpha^2+\left[1+\alpha^2+\left(\frac{\partial y}{\partial x'}\right)^2\alpha^2\right]^2}$ and also $\alpha > 0$.

The expression for the relative permittivity and permeability tensors becomes $\underline{\underline{\varepsilon'}} = \mathrm{diag}(\Lambda_2, \Lambda_2, \Lambda_3)$ and $\underline{\underline{\mu'}} = \mathrm{diag}(\Lambda_1, \Lambda_2, \Lambda_3)$. If we now multiply these expressions by $\alpha$, we obtain the reduced tensor components along the main optical axes, and importantly $\Lambda_3 = 1$. This means we now have only two varying tensor components for the permittivity, with the permeability tensor reducing to the identity. We further looked at a design for which $\Lambda_1 \sim \Lambda_2$.

The dispersion relation (15) reduces to:

$$k_x = \frac{\omega}{c}\sqrt{\frac{\Lambda_3\varepsilon_1(\Lambda_2\varepsilon_1 - \Lambda_1)}{\varepsilon_1^2 - \Lambda_1\Lambda_3}}. \quad (17)$$

It is clear from (17) that what matters to control the propagation of the surface plasmon is the vertical anisotropy i.e. the ratio between $\Lambda_1$ and $\Lambda_3$, whereas the transverse (magnetic) anisotropy encompassed in $\Lambda_2 = \mu_{yy2}$ has little to do with the invisibility carpet. However, controlling surface plasmons in such a way that they propagate around an obstacle on a structured surface (i.e. making a plasmonic invisibility cloak) further requires some transverse magnetic anisotropy, which is an additional challenge.

# Fundamental properties of SPPs

The penetration length of the SPP in both media depends on the dielectric constant and can be easily expressed as:

In metal:

$$z_{metal} = \frac{\lambda}{2\pi} \sqrt{\frac{\Re(\varepsilon_1) + \varepsilon_2}{\varepsilon_1^2}} . \tag{18}$$

In air:

$$z_{air} = \frac{\lambda}{2\pi} \sqrt{\frac{\Re(\varepsilon_1) + \varepsilon_2}{\varepsilon_2^2}} . \tag{19}$$

The propagation length of the SPP is given by:

$$L = \frac{c}{\omega} \left| \frac{\Re(\varepsilon_1) + \varepsilon_2}{\varepsilon_1 \varepsilon_2} \right|^{3/2} \frac{\Re(\varepsilon_1)^2}{I(\varepsilon_1)} . \tag{20}$$

In the case of the transformed material the propagation length $L_t$ is given by the following equations:

For the sake of clarity we define:

$$\varepsilon_1 = \varepsilon_r + \iota \varepsilon_i, \quad \varepsilon_{xx2} = \varepsilon_{x2}, \quad \varepsilon_{yy2} = \varepsilon_{y2}, \quad \varepsilon_{zz2} = \varepsilon_{z2} \tag{21}$$

$$X = \frac{\omega^2 \sqrt{\varepsilon_{z2}^2 \varepsilon_r^2 - 2\varepsilon_{z2} \varepsilon_r \varepsilon_i^2 + \varepsilon_i^4 + \varepsilon_i^2 \mu_{y2}^2 \varepsilon_r^2 - 2\varepsilon_i^2 \mu_{y2} \varepsilon_r \varepsilon_{x2} + \varepsilon_i^2 \varepsilon_{x2}^2}}{c^2 \sqrt{\varepsilon_r^4 + 2\varepsilon_i^2 \varepsilon_r^2 - 2\varepsilon_r^2 \varepsilon_{x2} \varepsilon_{z2} + \varepsilon_i^4 + 2\varepsilon_i^2 \varepsilon_{xx2} \varepsilon_{z2} + \varepsilon_{x2}^2 \varepsilon_{z2}^2}} \tag{22}$$

$$Y = \frac{\omega^2 \left( -\varepsilon_{z2}\varepsilon_r^3 + \varepsilon_{z2}\varepsilon_r\varepsilon_i^2 + \varepsilon_{z2}^2\varepsilon_r\varepsilon_{x2} + \varepsilon_i^2\varepsilon_r^2 - \varepsilon_i^4 - \varepsilon_i^2\varepsilon_{x2}\varepsilon_{z2} - 2\varepsilon_i^2\varepsilon_r^2\mu_{y2} + 2\varepsilon_r\varepsilon_i^2\varepsilon_{x2} \right)}{c^2 \left( \varepsilon_r^4 + 2\varepsilon_i^2\varepsilon_r^2 - 2\varepsilon_r^2\varepsilon_{x2}\varepsilon_{z2} + \varepsilon_i^4 + 2\varepsilon_i^2\varepsilon_{x2}\varepsilon_{z2} + \varepsilon_{x2}^2\varepsilon_{z2}^2 \right)} \quad (23)$$

So that:

$$L_t = 1/2 \frac{\sqrt{2}}{\sqrt{X+Y}} \quad (24)$$

In the experiment, at 800 nm, the propagation length is calculated to be around $45\,\mu m$ which is validated experimentally as shown in the following SEM image (Fig.5). We note that (24) indeed reduces to (20) provided that we assume that $\mu_{yy2} = 1$ and $\varepsilon_{xx2} = \varepsilon_{zz2} = \varepsilon_2$ in expressions (22) and (23), as it should.

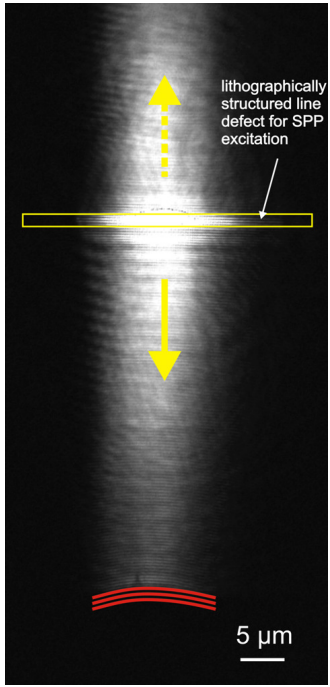

Figure 4. Experimental image of the leakage radiation of the SPPs interacting with a curved Bragg mirror. The SPPs are excited by slightly focussing the light at the lithographically structured defect line (marked by the yellow rectangular) and propagate under 90° at the suface away in both directions as indicated by the yellow arrows. The attenuation can be clearly observed and has been found to be similar to the calculations.

**Modelling of SPPs with finite elements**

In order to compute the total electromagnetic field for a SPP normally incident upon a three-dimensional carpet, we implemented the weak form of the scattering problem in the finite element package COMSOL using second order finite edge elements which behave nicely under geometric changes. Perfectly Matched Layers (PMLs), which can be seen as a stretch of coordinates, further enable us to model the unbounded three-dimensional domain. We choose the magnetic field $\mathbf{H}=(H_1, H_2, H_3)(x,y,z)$ as the unknown and therefore look for solutions of

$$\nabla \times \left( \underline{\underline{\varepsilon}}'^{-1} \nabla \times \mathbf{H} \right) - k_0^2 \mathbf{H} = \mathbf{0}, \tag{25}$$

where $k_0 = \omega\sqrt{\mu_0 \varepsilon_0} = \omega/c$ is the wavenumber, $c$ being the speed of light in vacuum, and $\underline{\underline{\varepsilon}}'$ is defined by:

$$\underline{\underline{\varepsilon}}' = \varepsilon_0 \mathbf{T}^{-1}, \quad \text{and} \quad \underline{\underline{\mu}}' = \mu_0 \mathbf{T}^{-1} \text{ where } \mathbf{T} = \mathbf{J}^T \mathbf{J}/\det(\mathbf{J}), \tag{26}$$

Also, $\mathbf{H} = \mathbf{H}_i + \mathbf{H}_d$, where $\mathbf{H}_i$ is the SPP incident field given by (2). The expression for $\mathbf{H}_i$ has been enforced on a flat surface on the leftmost part of the computational domain and $\mathbf{H}_d$ is the diffracted field which decreases inside the PMLs. We note that PMLs involve both an anisotropic heterogeneous medium with absorption above the interface $z=0$, which is fairly standard, but also an anisotropic heterogeneous metal with gain below the interface. The typical number of degrees of freedom used in our computations is around $5.10^6$ and computing time is approximatively 15 hours on a supercomputer with 256 Gb of RAM.

**Numerical study of the carpet from 650 nm to 900 nm**

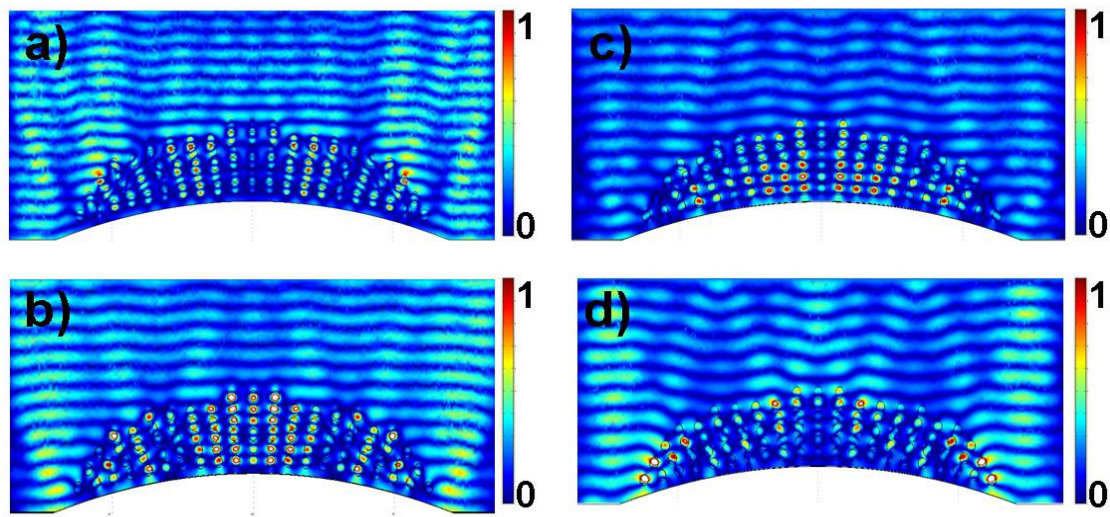

Figure 5. Numerical diffraction of a SPP incident from the top. Magnitude of the magnetic field is represented for different wavelengths a) 650 nm, b) 700 nm, c) 800 nm, b) 900 nm, respectively.